\documentclass{article}

\usepackage{spconf}
\usepackage{graphicx}
\usepackage{booktabs} 

\title{Improving Device Directedness Classification of Utterances with Semantic Lexical Features}
			    
\name{\parbox{\linewidth}{\centering Kellen Gillespie$^{\dagger}$, Ioannis C. Konstantakopoulos$^{\dagger}$, Xingzhi Guo$^{\star}$, Vishal Thanvantri Vasudevan$^{\dagger}$, \\ Abhinav Sethy$^{\dagger}$ }}
			\address{$^{\dagger}$ Amazon.com, Inc. \\
			    $^{\star}$ Stony Brook University }

\begin{document}

\maketitle

\begin{abstract}
User interactions with personal assistants like Alexa, Google Home and Siri are typically initiated by a wake term or wakeword. Several personal assistants feature ``follow-up'' modes that allow users to make additional interactions without the need of a wakeword. For the system to only respond when appropriate, and to ignore speech not intended for it, utterances must be classified as device-directed or non-device-directed. State of the art systems have largely used acoustic features for this task, while others have used only lexical features or have added LM-based lexical features. We propose a directedness classifier that combines semantic lexical features with a lightweight acoustic feature and show it is effective in classifying directedness. The mixed-domain lexical and acoustic feature model is able to achieve 14\% relative reduction of EER over a state of the art acoustic-only baseline model. Finally, we successfully apply transfer learning and semi-supervised learning to the model to improve accuracy even further.
\end{abstract}


\section{Introduction}

Personal assistants such as Alexa, Google Home, and Siri are becoming commonplace in modern households  \cite{sarikaya2017}. These systems respond to user commands prefixed by a wake command or \textit{wakeword}, e.g. \textit{``Alexa, what time is it?''} or \textit{``Hey Google, what's the weather?''}. This wakeword acts as signal that the utterance is intended for the assistant, in other words \textit{device-directed}.

To make consecutive commands more natural and conversational, personal assistants have added ``follow-up'' modes that allow users to give consecutive commands without repeated use of the wakeword. Recent examples include Baidu DuerOS's ``Full-Duplex'' feature, Google Home's ``Continued Conversation'' feature and Amazon Alexa's ``Follow-up Mode'' \cite{baidu, google, alexa}. These features allow for more natural conversations as shown in Table \ref{conversation}.

\begin{table}
\begin{center}
\begin{small}
\begin{sc}
\resizebox{\columnwidth}{!}{\begin{tabular}{ll}
\toprule
Standard Mode & Follow-up Mode \\
\midrule
Computer, what's the weather? & Computer, what's the weather? \\
(Device Responds) & (Device Responds) \\
Computer, what about tomorrow? & What about tomorrow? \\
(Device Responds) & (Device Responds) \\
Computer, thank you & Thank you \\
\bottomrule
\end{tabular}}
\end{sc}
\end{small}
\vskip -0.25cm
\caption{More Natural Conversation with Follow-up Modes}
\end{center}
\label{conversation}
\vskip -0.5cm
\end{table}

To achieve this behavior, the device's microphone is automatically re-opened after the device responds to the user. Users can then give device-directed follow-up commands without the need of a wakeword. However, if a user has no follow-up commands to give, the system could pick up speech not intended for it or \textit{non-device-directed} speech. Non-device-directed speech includes background speech, speech from media sources, and any other utterance not intended for the device. To alleviate this problem, and to avoid responding to non-device-directed utterances, virtual assistants employ directedness classifiers to separate utterances intended for the device from those that are not.

Directedness classification is an important task in many systems. Many such systems use acoustic features to determine directedness. In the personal assistant space, Mallidi et al.\ \cite{mallidi2018} leverage ASR decoder features as well as acoustic embeddings from the audio signal to perform directedness classification for one of the aforementioned follow-up modes. We use this work as the acoustic-only baseline in our evaluations.

Another system in the personal assistant space makes use of acoustic features and improves performance through the addition of an attention mechanism \cite{norouzian2019}. In related smart-environment applications, Reich et al.\ \cite{reich2011} use ASR decoder and prosodic features to perform directedness classification and Paek et al.\ \cite{paek2000} combine signal and user behavior characteristics to determine directedness in a continuous listening presentation domain. 

Other systems rely on language features to determine directedness \cite{lee2013}. In a simulated planetary exploration scenario, Dowding, et al.\ \cite{dowding2006} develop a system to help robotic members of mixed human robot teams determine if utterances were intended for them or other members. Shriberg et al.\ \cite{shriberg2012} add LM-based lexical features and contextual similarity features to supplement acoustic features in a Human-Human-Computer dialog environment.

Our proposed model will leverage signals from both acoustic and lexical domains. Other systems that have combined these domains have used LM-based or character-level lexical features \cite{mallidi2018, shriberg2012}. We propose a system to combine acoustic features with word-level semantic lexical features to improve directedness classification. We further improve the model with contextual features and attention mechanisms \cite{norouzian2019, shriberg2012}. To reduce the data labeling burden required by the model, we use transfer learning from a related dataset and experiment with semi-supervised learning techniques.

The paper is organized as follows. In Section \ref{data}, we look at the follow-up dataset to motivate the features and architectures to be evaluated. In Section \ref{models}, we describe our baseline feed-forward model as well as our proposed LSTM architectures. In Section \ref{experimentation}, we evaluate the discussed models and features and discuss the results. In Section \ref{semi-supervised}, we examine the effectiveness of a simple semi-supervised learning technique on our model training. In Section \ref{conclusion}, we discuss our overall findings and plans to improve our model. 

\section{Data Analysis} \label{data}

The primary feature of our proposed model is token-level semantic embeddings of the current turn utterance text derived from pre-trained word embeddings. Sentence embeddings can typically be derived by averaging token-level embeddings, but in this section we examine the data to identify additional features and architectures to develop a better model.

\subsection{Utterance-level Structure}

Many non-device-directed utterances in our data, due to factors such as containing multiple speakers or ASR errors, show a lack of sentence structure. Table \ref{structure} shows some examples of structured and non-structured utterances from the non-device-directed class.

\begin{table}[]
\begin{center}
\begin{small}
\begin{sc}
\resizebox{\columnwidth}{!}{\begin{tabular}{ll p{\columnwidth}}
\toprule
Structured & Unstructured \\
\midrule
did you reorder your pills		& it's it's just flashing yellow \\
I don't know what she just order		& can you hey computer \\
well you got four more hours		& break at a bigger \\
mom what did you say		& tell us moving \\
what are you doing		& weather talking about hal \\
\bottomrule
\end{tabular}}
\end{sc}
\end{small}
\end{center}
\vskip -0.5cm
\caption{Structured and Unstructured Utterances Misidentified as Device-directed}
\label{structure}
\end{table}

A model that accounts for word order, and thus indirectly models grammatical structure, can be expected to learn to reject such unstructured utterances as non-device-directed. For this reason, we expect a recurrent model such as an LSTM will outperform a baseline DNN model on this task.

\subsection{Contextually-Relevant Utterances} \label{context}

While the current turn utterance can normally provide most of the information needed to perform directedness classification, there is a subset of our dataset that contains contextually-relevant utterances. We give a few examples in Table \ref{context_examples}.

\begin{table}[h]
\begin{center}
\begin{small}
\begin{sc}
\resizebox{\columnwidth}{!}{\begin{tabular}{ll p{\columnwidth}}
\toprule
Previous Turn & Current Turn \\
\midrule
computer add bananas to my shopping list	& add chicken \\
computer what's the weather for today		& and for tomorrow \\
computer who is adele?					& play \\
computer are you unique?					& are you one of a kind? \\
\bottomrule
\end{tabular}}
\end{sc}
\end{small}
\end{center}
\vskip -0.5cm
\caption{Utterances with Relevant Previous Turns}
\label{context_examples}
\end{table}

Looking solely at the current turn, these examples may be difficult to classify. With the context of the previous turn, however, they more clearly become interpretable as device-directed utterances. For this reason, we experiment with incorporating the previous turn text into the features to aid in classification.

\subsection{Textual Ambiguity} \label{ambiguity}

While we expect semantic lexical features to aid in directedness classification, they do not on their own solve the problem of textual ambiguity present in our data. Many utterances belong to both classes in our dataset, in fact 29.7\% of the dataset consists of utterances belonging to both classes. We highlight some examples below in Table 4, showing the most frequent occurring utterances belonging to both classes. For each utterance, we look at class frequencies to determine prior probabilities of each belonging to each class.

\begin{table}[h]
\begin{center}
\begin{small}
\begin{sc}
{\begin{tabular}{lll}
\toprule
Utterance & p(DD) & p(NDD) \\
\midrule
thank you & 93.2 & 6.8 \\
stop & 44.8 & 55.2 \\
okay & 45.8 & 55.2 \\
cancel & 70.7 & 29.3 \\
what & 32.6 & 67.4 \\
next & 62.8 & 37.2 \\
good night & 61.6 & 38.4 \\
play & 36.8 & 63.2 \\
\bottomrule
\end{tabular}}
\end{sc}
\end{small}
\caption{Textual ambiguity of utterances common to device-directed (DD) and non-device-directed (NDD) classes.}
\end{center}
\label{amb_examples}
\vskip -0.5cm
\end{table}

It's clear that for many frequent utterances, text-only input will be insufficient for distinguishing between classes. For this reason, we will fuse the lexical features with acoustic features to assist in separating these ambiguous cases. Our acoustic feature will be token-level confidence scores from an upstream ASR decoding component \cite{gales2008}. We also expect a subset of these cases to be assisted by the contextual features described in Section \ref{context}.

\section{Candidate Models} \label{models}

\subsection{Feature Representation}

\begin{figure}[]
\begin{center}
\centerline{\includegraphics[width=\columnwidth]{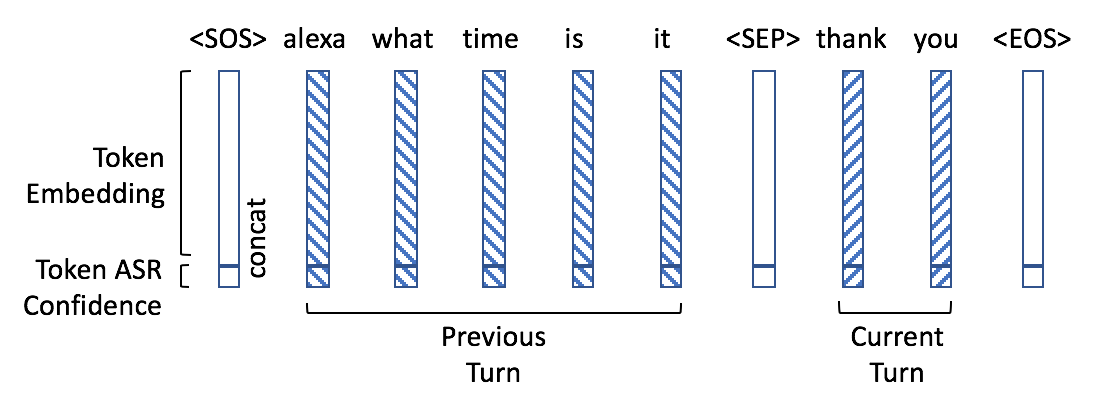}}
\vskip -0.25cm
\caption{Input Feature Representations}
\label{features}
\end{center}
\vskip -1cm
\end{figure}

We experiment with three features for our candidate models: the ASR-produced text of the current turn, the text of the previous turn, and token-level ASR confidence scores. The turns are separated by special tokens similar to those used in other systems \cite{devlin2018}. We concatenate the lexical features, the word embeddings of each token, with their corresponding acoustic feature in the form of the ASR confidence scores. For pre-trained embeddings we use fastText wiki news word vectors \cite{mikolov2018} with an embedding size of 300. The only token embedding we specifically train for our data is an out-of-vocabulary or OOV token to better represent the OOV tendencies of personal assistant commands, while all other embeddings are left as-is. Due to the trained OOV token we do not make use of subword vectors in fastText. Figure \ref{features} shows the overall input representation of our features.

\subsection{AVG-DNN Model}

The non-recurrent proposed model is a fully-connected feed-forward network using a word average sentence representation of the utterance as input. The sentence embedding is simply the average of the token features shown in Figure \ref{features}. The DNN has 3 layers: an input layer, dense hidden layer of size 150, and an output layer.

\subsection{LSTM Model}

The LSTM model \cite{hochreiter1997} makes use of sequential information in the input features. For this reason, the token-level features in Figure \ref{features} are used as-is, with each token representing an input frame. The LSTM has 3 layers of size 150 followed by a single dense layer and an output layer. Bidirectional LSTM (BLSTM) models were tested but did not improve over the single direction versions.

\subsection{LSTM+Attention Model}

Attention mechanisms are a common addition to LSTM architectures in neural translation and even directedness classification tasks \cite{bahdanau2015, norouzian2019}. Instead of using the final LSTM hidden state as the sentence embedding, the Attention model performs a soft selection over the hidden states of the entire sequence to generate a sentence embedding. The attention mechanism is a simple affine layer activated with a tanh function before being passed to a softmax selection layer.

\section{Experimentation} \label{experimentation}

Our dataset consists of follow-up utterances with ground truth annotations of device-directed or non-device-directed. The dataset contains roughly 240K utterances and is spilt into train, dev, and test partitions with 200K, 20K, and 20K utterances, respectively. The class split is roughly 5:1 in favor of the device-directed class in each partition, as non-device-directed speech is far less frequent.

\subsection{Transfer Learning}

While the dataset described above is not small, there is far more data available for first-turn interactions. We pre-train all models on a very large (2M) ASR false-wake dataset that contains first-turn device-directed speech and non-device-directed false-wake speech. False-wake speech are instances where the device mistakenly heard the wakeword. While first-turn and false-wake data is not exactly like follow-up mode data, the tasks are related enough that we consistently see improvement when pre-training on this data. We then fine tune the models on the proper follow-up dataset.

\subsection{Model Training}

The models were trained using SGD with the cross entropy loss function. We allowed for a decaying learning rate, and chose maximum and minimum learning rates using LR range tests \cite{smith2015}. For transfer learning, we pre-train with higher learning rates and fine-tune with smaller learning rates.

\subsection{Model Evaluation}

Table 5 shows the initial evaluation results for the candidate models, as well as an ablation test for our best model. We compare with the acoustic-only baseline model \cite{mallidi2018}, but due to implementation and feature differences this model was trained once on a combination of our pre-trained and follow-up datasets. The DNN model is compared with and without the previous turn feature, as its average sentence embedding is unable to distinguish between turns. For the best model, we perform an ablation study to measure feature contributions and compare accuracy with and without transfer learning. Since our classes are imbalanced, we utilize the equal error rate (EER) metric.

\begin{table}[h]
\begin{center}
\begin{small}
\begin{sc}
{\begin{tabular}{lccc}
\toprule
Model & Features & EER \(\%\) \\
\midrule
Acoustic-only Baseline \cite{mallidi2018} $^\dagger$ & -- & 10.6 \\
AVG-DNN		& c,t		& 19.2  \\
AVG-DNN		& c,p,t	& 19.4  \\
LSTM		& c,p,t	&  9.2  \\
LSTM+Attn	& c,p,t 	&  \textbf{9.1}  \\
\midrule
LSTM+Attn	& c,p,t 	&  9.1  \\
			& -c		&  13.7 \\
			& -p		&  9.5  \\
			& -t 		&  14.8  \\
LSTM+Attn-NoTL	& c,p,t 	&  10.6 \\
\bottomrule
\end{tabular}}
\end{sc}
\end{small}
\caption{Model evaluation with feature ablation for the best model. \textbf{C}: current utterance text, \textbf{P}: previous utterance text, \textbf{T}: token-level ASR confidences, $^\dagger$: trained in a different manner}
\end{center}
\label{eval_results}
\vskip -0.5cm
\end{table}

The LSTM is able to outperform the average embedding model by a significant margin. The LSTM model with attention performs even better, though the improvement is much smaller. The best model with all features is able to reduce EER by 14\% relative over the baseline model based on acoustic features.

The ablation study with the best model shows that all features add complementary predictive power to the model, albeit at different magnitudes. The token-level ASR confidences are the most important feature, quickly followed by the current turn text. The previous turn is less critical, likely due to the fact that only a subset of the dataset contains contextually-relevant utterances. Finally, our best model reduces EER by 14\% relative over the same one trained without transfer learning \textsl{LSTM+Attn-NoTL}.

\section{Semi-Supervised Learning} \label{semi-supervised}

Labeling data is time consuming and costly. Semi-supervised learning approaches allow a model trained on labeled data to continue training on unlabeled data. Many approaches exist, but for simplicity we first investigate a self-teaching methodology. We use two models to aid in labeling like a co-training approach \cite{blum1998}. In this section we describe in detail our method and the results.

\subsection{Method}

We employ the self-teaching semi-supervised learning approach. In this method, first a model is trained in a standard supervised approach with labeled data $L$. Next, the trained model performs inference on a large set of unlabeled data $U$. The unlabeled cases where the model is highly confident $U_L$ are labeled, removed from the unlabeled data and folded into training set. The model is then retrained with the new dataset $L + U_L$ and the process is repeated until stopping criteria is met.

We expand on the method by combining the scores of two distinct models, our proposed model and the acoustic-only baseline model, to reduce risk by labeling only those cases where the models have agreement. We apply a nonlinear transformation to the acoustic model score to lower its contribution to the overall score, notably in high-confusion posterior ranges.

\subsection{Evaluation}

Our unlabeled dataset $U$ consists of 500K utterances. We select the LSTM without contextual features or attention; the unlabeled data does not contain previous turns and the attention model, likely due to hyperparameter issues, did not achieve good results during semi-supervised training. At each iteration we label the highest 1\% of scores as DD and the lowest 0.2\% of scores as NDD, to maintain the original class priors of the labeled dataset.

The stopping criteria is met once dev loss is no longer improving, in this case after 20 passes of semi-supervised training. We choose pass 15, the point of lowest dev loss, as our selected model. We evaluate this model, as well as several other versions, on the original test set and compare to the original model. Table \ref{semi_results} shows test set EER and dev set loss over select passes during semi-supervised training. Test EER improves 5\% relative due to semi-supervised learning, and our stopping point based on the dev loss achieves the best EER.

\begin{table}[]
\begin{center}
\begin{small}
\begin{sc}
{\begin{tabular}{ccc}
\toprule
Semi-supervised Pass & Test EER \(\%\) & Dev Loss \\
\midrule
0		& 10.4 &  0.176 \\
5		& 10.1  & 0.175 \\
10		& 10.0  & 0.173\\
15 [stopping point]	&  \textbf{9.9}  & 0.170 \\
20		& 9.9  & 0.171\\
\bottomrule
\end{tabular}}
\end{sc}
\end{small}
\vskip -0.125cm
\caption{Semi-supervised model evaluation}
\label{semi_results}
\end{center}
\vskip -0.5cm
\end{table}

\section{Conclusions \& Future Work} \label{conclusion}

We have proposed a directedness classifier for spoken utterances based on acoustic and semantic lexical features. We have shown token-level semantic information, informed from word embeddings, can classify utterances with reasonable accuracy. Supplementing these lexical features with a lightweight acoustic feature and contextual information improves accuracy even further. Our results show that a recurrent model, such as an LSTM with attention, can utilize these features to reduce EER by 14\% relative over an acoustic-only baseline model. Finally, we show that transfer learning and semi-supervised techniques can further improve the model without increasing the burden of data labeling.

Our future work will be along several dimensions. First, we plan to investigate more model architectures, such as CNN-LSTMs, FLSTMs, and Transformer-style models. For semi-supervised learning, we plan to experiment with much larger unlabeled datasets and more advanced graph-based techniques for exploration during training. We further plan to leverage contextual features and attention in our semi-supervised experimentation.

\clearpage

\bibliographystyle{IEEEbib}
\bibliography{dd} 

\end{document}